\documentclass[12pt]{iopart}
\usepackage{microtype}
\usepackage{graphicx}
\usepackage{subfigure}
\usepackage{booktabs} % for professional tables

\usepackage{hyperref}
\usepackage[capitalize]{cleveref}
\Crefname{equation}{Equation}{Equations}
\crefname{equation}{Eq.}{Eqs.}
\Crefname{figure}{Figure}{Figures}
\crefname{figure}{Fig.}{Figs.}
\Crefname{table}{Table}{Tables}
\crefname{table}{Tab.}{Tabs.}
\Crefname{section}{Section}{Sections}
\crefname{section}{Sec.}{Secs.}
\Crefname{subsection}{Subsection}{Subsections}

\begin{document}

\title[Robust SBI in Cosmology with BNNs]{Robust Simulation-Based Inference in Cosmology \\
with Bayesian Neural Networks}

\author{Pablo Lemos\textsuperscript{1, 2}, Miles Cranmer\textsuperscript{3}, Muntazir Abidi\textsuperscript{4}, ChangHoon Hahn\textsuperscript{3}, Michael Eickenberg\textsuperscript{5}, Elena Massara\textsuperscript{6, 7}, David Yallup\textsuperscript{8} and Shirley Ho\textsuperscript{5, 3, 9, 10}}

\address{$^1$ Department of Physics and Astronomy, University of Sussex, Sussex House, Falmer, Brighton, BN1 9RH, UK}
\address{$^2$ Department of Physics and Astronomy, University College London, Gower Street, London, WC1E 6BT, UK}
\ead{p.lemos@sussex.ac.uk}

\address{$^3$ Department of Astrophysical Science, Princeton University, Peyton Hall, Princeton NJ 08544, USA}

\address{$^4$ D\'epartement de Physique Th\'eorique, Universit\'e de Gen\`eve, 24 quai Ernest Ansermet, 1211 Gen\`eve 4, Switzerland}

\address{$^5$ Flatiron Institute Center for Computational Mathematics, 162 5th Ave, 3rd floor, New York, NY 10010, USA}

\address{$^6$ Waterloo Centre for Astrophysics, University of Waterloo, 200 University Ave W, Waterloo, ON N2L 3G1, Canada}

\address{$^7$ Department of Physics and Astronomy, University of Waterloo, Waterloo, ON N2L 3G1, Canada}

\address{$^8$ Kavli Institute for Cosmology, Cavendish Laboratory, Madingley Road, Cambridge, CB3 0HA, UK}

\address{$^9$ Center for Cosmology and Particle Physics, Department of Physics, New York University, New York, NY 10003, USA}

\address{${10}$ Department of Physics, Carnegie Mellon University, Pittsburgh, PA 15213, USA}

\vspace{10pt}
\begin{indented}
\item[]August 2017
\end{indented}
\begin{abstract}
    Simulation-based inference (SBI) is rapidly establishing itself as a standard machine learning technique for analyzing data in cosmological surveys. Despite continual improvements to the quality of density estimation by learned models, applications of such techniques to real data are entirely reliant on the generalization power of neural networks far outside the training distribution, which is mostly unconstrained. Due to the imperfections in scientist-created simulations, and the large computational expense of generating all possible parameter combinations, SBI methods in cosmology are vulnerable to such generalization issues. Here, we discuss the effects of both issues, and show how using a Bayesian neural network framework for training SBI can mitigate biases, and result in more reliable inference outside the training set. We introduce {\tt cosmoSWAG}, the first application of Stochastic Weight Averaging to cosmology, and apply it to SBI trained for inference on the cosmic microwave background.
    \end{abstract}
    
    %\vspace{-0.8cm}
    
    \section{Introduction}
    \label{sec:introduction}
    %\vspace{-0.1cm}
    
    We are entering a new era for cosmology. Machine Learning applications to cosmology allow for the analysis of large datasets, and the exploration of new models and phenomena~\cite{kangal2019machine, ntampaka2019role, escamilla2020deep, tilaver2021deep, salti2021evolution, dvorkin2022machine}
     Traditionally, the field has relied on likelihood-based methods, in which we compress our data into summary statistics, for which we can make theoretical predictions and build likelihood functions.
    However, with the development of practical machine learning tools for high-dimensional data over the last decade, it is now possible to perform cosmological analysis even for intractable likelihoods. Instead of a likelihood, we can use simulations of observables to perform parameter inference, and model comparison. This technique is often called Likelihood-Free Inference, 
    approximate Bayesian computation (ABC)~\cite{csillery2010approximate, beaumont2010approximate, sunnaaker2013approximate}, implicit-likelihood inference (ILI) or simulation-based inference (SBI)~\cite{thomas2016likelihood}. We will adopt the latter term in the remainder of this work. SBI allows us to perform parameter inference and model comparison, even in situations where the likelihood is intractable, such as field-level inference~\cite{leclercq2021accuracy}. 
    
    Multiple SBI methods have been developed in recent years, but particularly relevant to cosmology is Density Estimation Likelihood-Free Inference (DELFI, also known as neural posterior estimation)~\cite{bonassi2011bayesian, fan2013approximate, papamakarios2016fast, lueckmann2017flexible, lemos2021sum}, which uses a density estimator to estimate the likelihood\footnote{Different variations of the method estimate the likelihood, and then multiply by the prior, or estimate the posterior directly.}. This method has multiple advantages: it uses all available simulations and estimates the full-dimensional posterior distributions, not just marginalised posteriors. 
    However, practical applications of DELFI to cosmology often encounter two issues~\cite{cranmer2020frontier}: The first one is the limited number of available simulations. To circumvent the curse of dimensionality, the original DELFI method proposes using a step of massive data compression, which reduces the dimensionality of the data to the dimensionality of the parameter space. This facilitates the task of density estimation. Proposed data compression methods include MOPED~\cite{heavens2017massive} and Information Maximizing Neural Networks (IMNNs)~\cite{charnock2018automatic, makinen2021lossless}. These methods, however, rely on either a covariance matrix for the data errors or the ability to generate a large number of simulations to estimate a covariance. When none of these conditions are met, other data compression methods have to be used, which will generally lead to a lossy compression -- meaning it does not retain all information about the parameters -- and a loss of accuracy. This will be the case if we intend to apply DELFI to an existing suite of simulations, such as the QUIJOTE~\cite{villaescusa2020quijote} and CAMELS~\cite{villaescusa2021camels} simulations.
    
    The second issue of practical applications of DELFI, and SBI in general, is difficulty simulating realistic observations. It does not matter how good the performance of our SBI algorithm is if we have failed to generate simulations that model all systematic effects and observational errors. In interesting examples, it is impossible to model everything. Therefore, we try to get as close as possible. But we need to deal with the fact that our simulations are likely to be imperfect. Furthermore, most SBI methods, including DELFI, have no way of informing us whether the observations we are trying to analyse are different from our simulations. How do we then interpret a surprising result coming from an SBI analysis? As a true scientific discovery, or a failure to generate realistic enough simulations? In this work, we present a way to mitigate this effect: We propose using Bayesian neural networks (BNNs) in our SBI analysis. BNNs are well known to provide better generalization to observations that have not been used during training~\cite{kononenko1989bayesian, mackay1995bayesian, gal2016dropout, 2022arXiv220511151Y}. We also expected BNNs to account for some of the epistemic uncertainty introduced in the neural network training. Therefore, in the presence of unknown systematics, BNNs will give us larger errors, instead of biased posteriors. With this goal in mind, in this work, we introduce {\tt cosmoSWAG}, %\footnote{
    %    The code will be made available upon acceptance of the paper.
    %} 
    the first application of stochastic weight averaging (SWA)~\cite{maddox2019simple, wilson2020bayesian} to cosmology.\footnote{The code is available at \url{https://github.com/Pablo-Lemos/cosmoSWAG}.}
    SWAG (SWA Gaussian) was previously used in astronomy \cite{cranmerBayesianNeuralNetwork2021} to accurately predict planetary instability of five-planet systems, despite only training on three-planet systems.
    While other methods exist to perform approximate marginalisation over neural network parameters, such as MC dropout~\cite{gal2016dropout, gal2017concrete} or Variational Inference~\cite{graves2011practical, kingma2013auto}, SWAG has been show to perform better over a variety of tasks~\cite{maddox2019simple}.
    
    The goal of this paper is to study how we can maximize the accuracy of a DELFI analysis, for a fixed suite of simulations, and in the case in which running more simulations is not possible. This is the situation we find ourselves in if we want to perform a DELFI analysis with existing data, in situations where simulations are costly. 
    
    \section{Simulator}
    \label{sec:simulator}
    
    \begin{figure}[t]
        \begin{center}
        \centerline{\includegraphics[width=0.95\columnwidth]{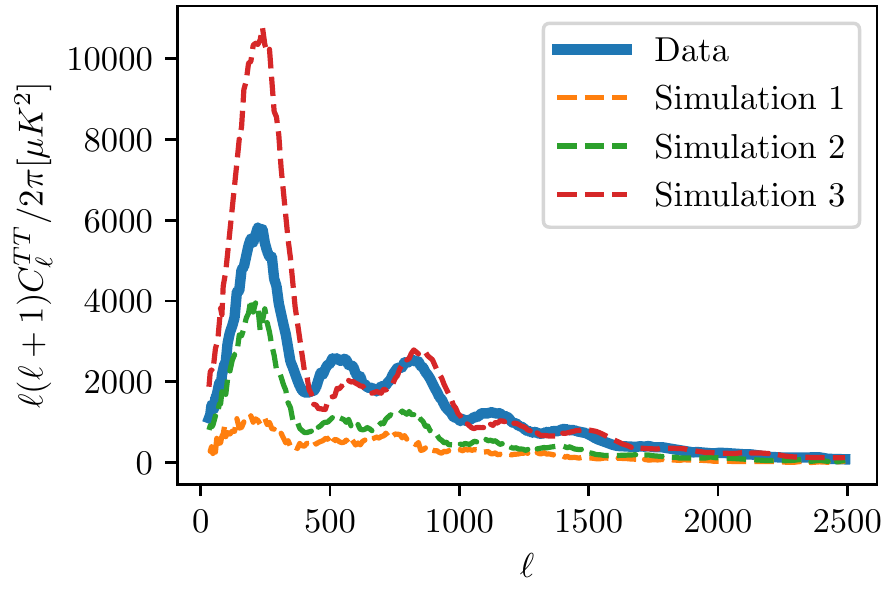}}
        \caption{Three example simulations, chosen from three random parameter draws from the prior~\cref{tab:prior}, and true data as observed by {\it Planck} in blue.}
        \label{fig:simulations}
        \end{center}
    \end{figure}

    To set up a realistic cosmological analysis, that we can apply DELFI to, we choose to use simulations of the Cosmic Microwave Background (CMB) power spectrum. The main reason to do this is that this is a problem where it is easy, and computationally cheap, to generate a suite of simulations; and that this is a problem where we can actually write down a likelihood and perform a likelihood-based analysis. Therefore, by using this simulator, we can compare our obtained posterior distributions to the ones we should obtain. \cite{cole2021fast} already used the CMB to test the performance of an SBI model. 
    
    Our approach is therefore the following: 
    
    1. We use {\tt CAMB}~\cite{Lewis:1999bs, Lewis:2002ah, Howlett:2012mh} to generate a suite of $100.000$ CMB power spectra\footnote{A previous version of this paper used a suite of $10.000$. The effect of varying the number of simulations is explored in~\ref{sec:number}.}. We use $\ell_{\rm max} = 2500$, and use only, the power spectrum of temperature anisotropies. 
    
    2. We then use {\it Planck} 2018 TT~\cite{aghanim2020planck} native likelihood used in the code {\tt cobaya}~\cite{torrado2021bcobaya}, to convert this power spectrum into a binned power spectrum, at 215 multipole bins. 
    
    3. We use the {\it Planck} TT data and covariance matrix in the same likelihood, as our observed data and error model, respectively. One advantage of this likelihood is that it uses only multipoles $\ell > 30$, and approximates the error model at those scales by a normal distribution. 
    
    We show some example simulations, as well as the true observation in~\cref{fig:simulations}. Our simulations are drawn from a uniform prior, shown in~\ref{sec:prior}.
    
    \section{Analysis}
    \label{sec:analysis}
    
    \begin{figure}[ht!]
        \begin{center}
        \centerline{\includegraphics[width=0.49\columnwidth]{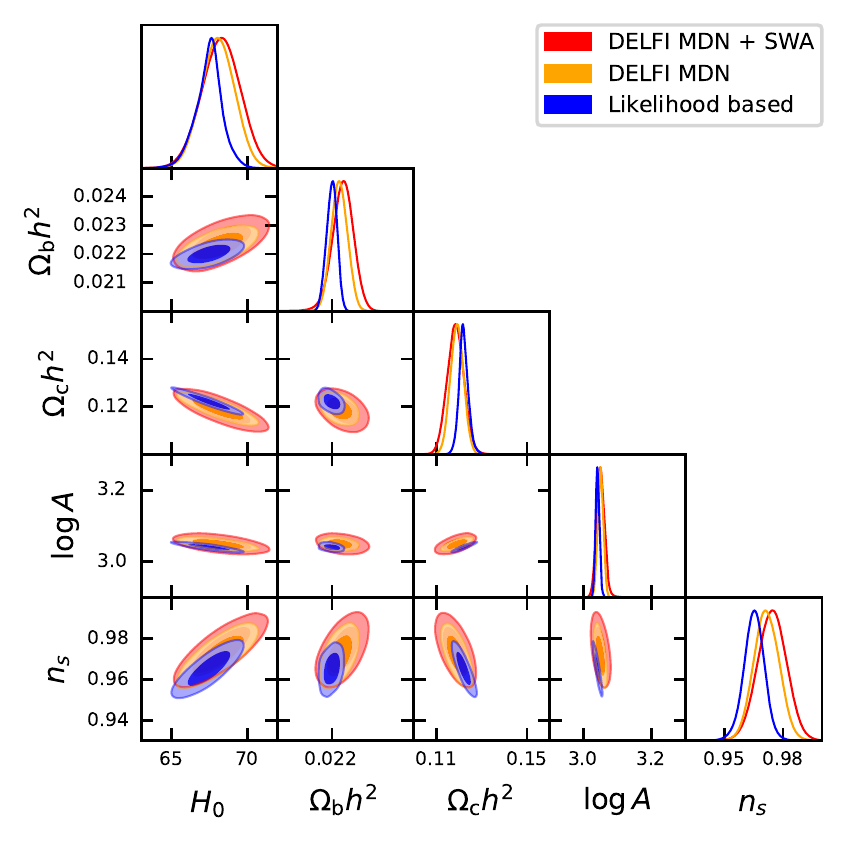}
        \includegraphics[width=0.49\columnwidth]{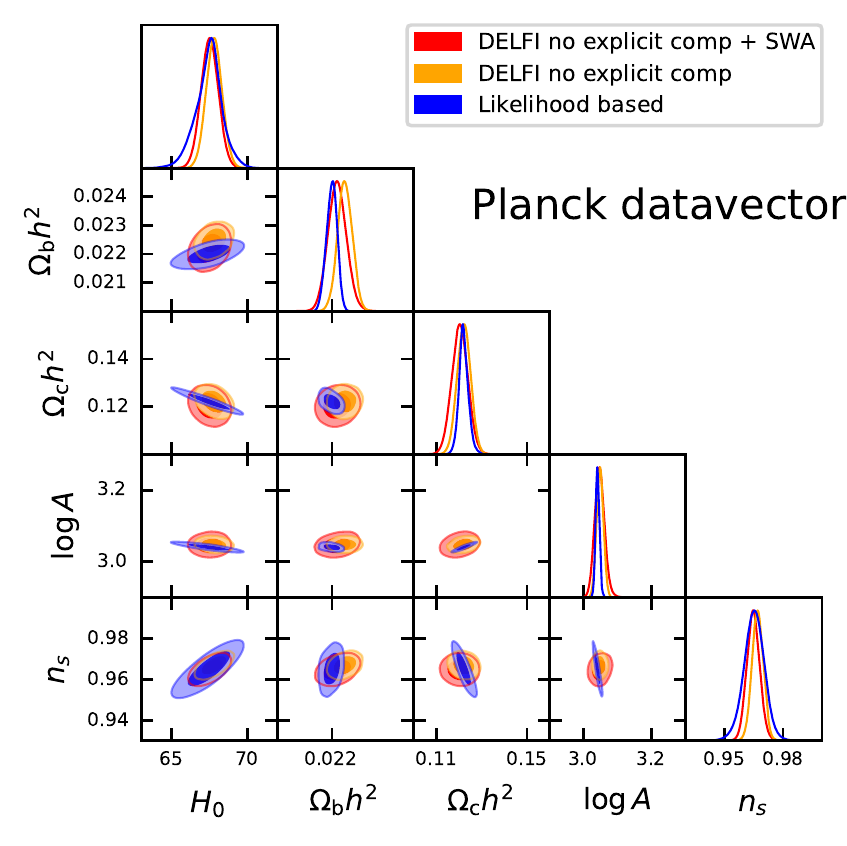}}
        \caption{One and two-dimensional marginalised posterior distributions for the different variations of DELFI presented in~\cref{sec:analysis} using {\it Planck} data. The left plot shows the results for DELFI without explicit compression, using an MDN, with weight-marginalised results in red, and non weight-marginalised in orange. In blue, we show the likelihood-based analysis. The right plot shows the same for DELFI with compression through a regression network. The contours represent the 68 and 95 \% confidence levels. This plot was generated using {\tt GetDist}~\cite{Lewis:2019xzd}}
        \label{fig:results}
        \end{center}
    \end{figure}
    
    \subsection{DELFI with massive data compression}
    \label{sec:delfi_compression}
    
    To perform parameter inference using our CMB simulations,
    we start by doing the DELFI analysis we would under ideal circumstances, as described in~\cite{alsing2018massive, alsing2019fast}. In this analysis, we start with a step of massive data compression that reduces the dimensionality of the data to the dimensionality of the parameter space. We can ensure this compression is lossless when data is abundant, e.g. in a situation when we can quickly generate large numbers of new simulations, through algorithms such as MOPED and IMNN. However, we want to test how well we can perform using only a fixed set of simulations, to simulate a realistic scenario. In that case, the
    neural network compression will be inaccurate and hence it may lose information about the parameters. We use a regression network, also known as a neural compressor; which is just a neural network that tries to predict parameter values from data. This compresses the data into the dimensionality of the parameter space, but that compression can be imperfect. We use a neural network with 6 hidden layers, each containing 128 neurons. We use rectified linear unit (ReLU)~\cite{agarap2018deep} activation functions, and $L_2$ regularization of the weights, with a regularization factor $0.1$. Our loss is the mean squared error. During training, we add noise to each input according to the noise model described in~\cref{sec:simulator}.
    
    We then use the predictions of this neural network as the compressed data in our DELFI analysis. We use a masked autoregressive flow (MAF)~\cite{papamakarios2017masked} as a density estimator, containing a stack of 5 masked autoencoders~\cite{germain2015made}, each containing two hidden layers with 30 neurons each. We do this using the {\tt pyDELFI} package available at \url{https://github.com/justinalsing/pydelfi}.
    
    \subsection{DELFI without explicit data compression}
    \label{sec:delfi_no_compression}
    
    Given that we expect the compression to be lossy, it is natural to ask ourselves the question: What about using no explicit data compression, and performing density estimation directly on the data? After all, density estimation techniques such as normalizing flows have been applied successfully to high dimensional data, such as images~\cite{helminger2020lossy}. Therefore, we try to perform our DELFI analysis directly from the data. 
    
    We use Mixture Density Networks (MDN)~\cite{bishop1994mixture} for density estimation, instead of MAFs. The reason for this is that we want to compare the results of this section, to the results of the following section using {\tt cosmoSWAG}, and at present time {\tt cosmoSWAG} does not support MAFs. %We did compare the results of this section using MDNs and MAFs and found that MAF gets slightly tighter contours, but the differences are not large enough to affect our conclusions. 
    
    Therefore, we use a neural network with the same structure as the one used in~\cref{sec:delfi_compression}, but with a different number of outputs, as described in~\ref{sec:MDN}. 
    
    \begin{figure}[ht!]
        \begin{center}
        \centerline{\includegraphics[width=0.49\columnwidth]{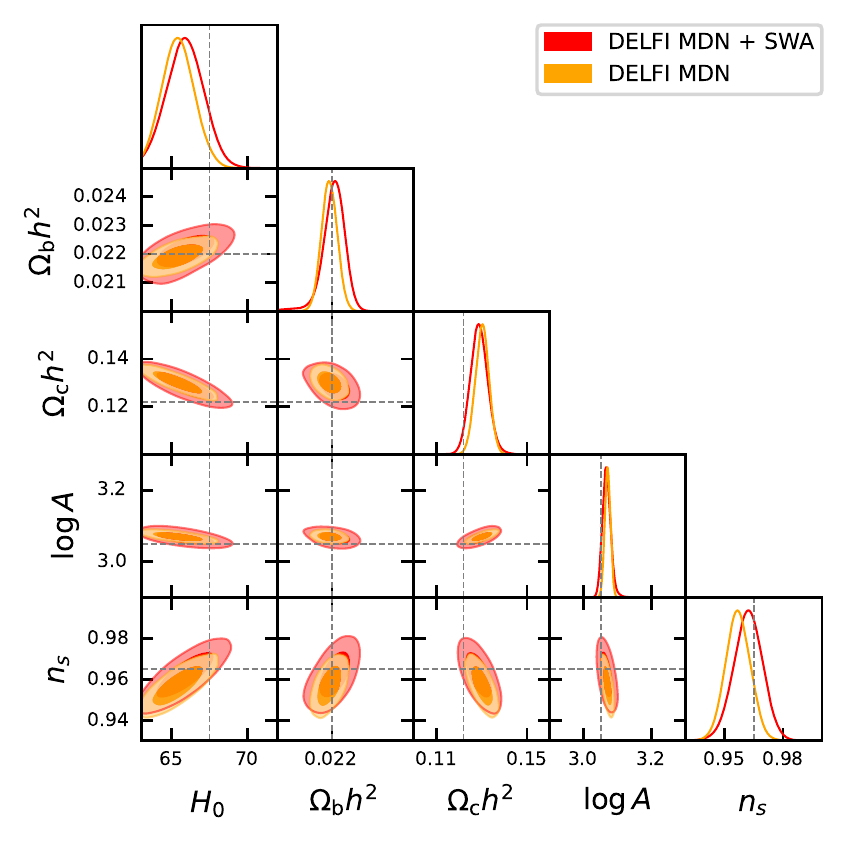}
        \includegraphics[width=0.49\columnwidth]{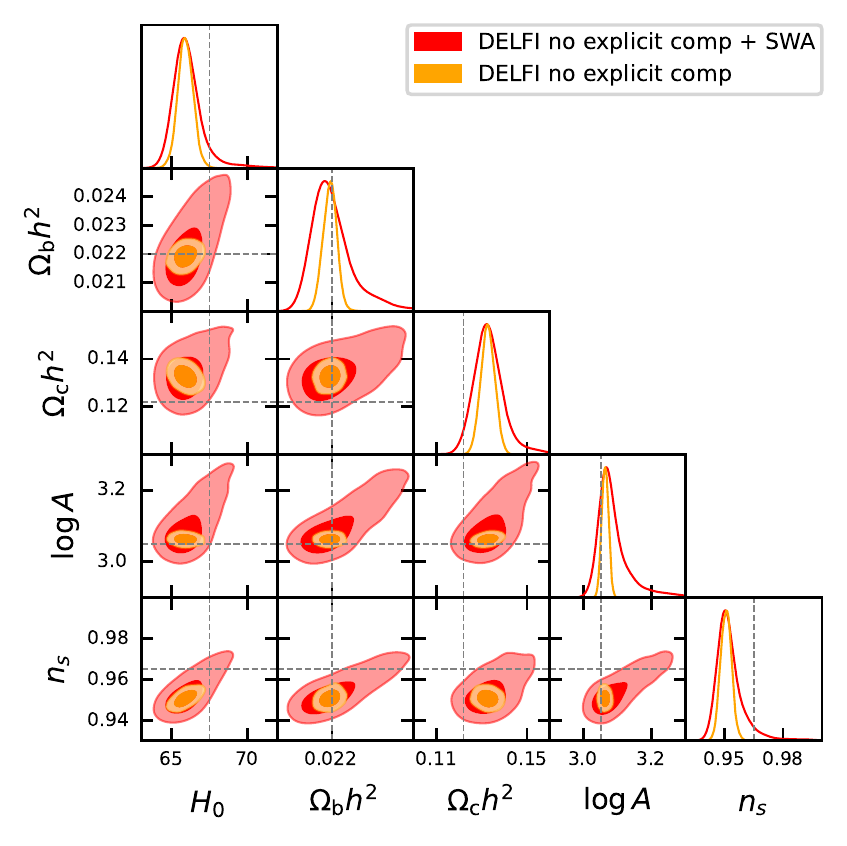}}
        \caption{
            The same as~\cref{fig:results} using a synthetic data vector with added noise, as described in~\cref{sec:generalization}.
            }
        \label{fig:results_noisy}
        \end{center}
    \end{figure}
    
    \subsection{DELFI without explicit data compression and with weight marginalisation}
    \label{sec:delfi_no_compression_marg}
    
    Next, we repeat the analysis of DELFI with an MDN, but applying SWA to the neural network. The basic idea is, starting from a pre-trained set of neural network weights, to perform stochastic gradient descent with a constant large learning rate. We average the weights as the model is trained, and use the evolving weight values to approximate a mean and covariance matrix for the neural network weights. While this assumes that the posteriors on the weights are Gaussian, the method provides an estimate of the weight uncertainty, and therefore the uncertainty of the predictions. More moments could be computed to characterize the posterior in more detail and assess the validity of the Gaussian assumption. Furthermore, when estimating the covariance matrix of the neural network weights, we use a tunable `scale' hyperparameter. The reason for this hyperparameter is that the covariance matrix estimated by SWA will depend on the learning rate. While for an optimal learning rate~\cite{mandt2017stochastic}, the scale parameter should be set to $0.5$, in practice it is possible to use the scale hyperparameter to rescale the covariance, and therefore the posterior width. In this work, we use the validation set to find the optimal value of the scale hyperparameter, as explained in~\cref{sec:scale}. A more detailed description of the method can be found in~\cite{maddox2019simple}. 
    
    BNNs provide two important advantages over traditional neural networks. Robustness to overfitting~\cite{hernandez2015probabilistic} and generalization properties~\cite{wilson2020bayesian}. Robustness to overfitting means that we are less likely to get biased posteriors. More importantly, the generalization properties mean that our SBI algorithm should perform better when our observed data does not perfectly match the simulations, either because of systematics or observational effects in the data that are not present in the simulations or because the theoretical model we are using to simulate is incorrect. We test this using our simulator in the following section.
    
    \subsection{DELFI with massive data compression and with weight marginalisation}
    \label{sec:delfi_no_compression_marg}
    
    Finally, we can repeat the DELFI analysis with explicit massive compression through a regression network, adding marginalisation to the neural network weights. 
    We first compress the data with a regression network, and the apply a MDN on the compressed data. We use weight marginalization on both the compression and the MDN.
    
    \section{Results}
    \label{sec:results}
    
    \subsection{Comparison with likelihood-based analysis}
    \label{sec:likelihood_analysis}
    
    The results of applying all four versions of our DELFI analysis are shown in~\cref{fig:results}. 
    We first focus on the left panel, using no explicit compression. We see that the DELFI posteriors do correctly capture the degeneracies of the likelihood, as the ellipses are `tilted' in the same way as the real ones. 
    
    The size of both DELFI posteriors is larger than the likelihood-based one. This is caused by the fact that we are using a limited number of simulations, and due to the added epistemic uncertainty of having to estimate a likelihood from simulations.
    When we include weight marginalisation with SWA, the size of the contours increases and improves the agreement with the expected result. 
    These results are further confirmed by repeating the analysis on several validation simulations, as shown in~\ref{sec:validation_default}: We get slightly underconfident results with DELFI, even before weight marginalisation, meaning we can trust the posteriors. 
    
    When we instead use data compression with a neural compressor, we see that we obtain slightly smaller contours.
    Weight marginalisation in this case leads to a small increase in the contours. 
    %Again, the fact that all our DELFI posteriors are larger than the likelihood-based posterior, is a consequence of our limited number of simulations. In general, when our number of simulations is limited, our SBI analysis will lose some of its potential constraining power. 
    In both cases, we notice that we no longer fully recover the degeneracy directions, as we did when we did not use a massive data compression step. This is likely due to some loss of information in the regression network. 
    Our validation test (shown in~\ref{sec:validation_default}) shows good results for this case, even before weight marginalization.
    
    \subsection{Generalization}
    \label{sec:generalization}
    
    In this section, we aim to test how our system behaves in the presence of unknown systematics or observational effects in the data, that are not present in the simulations. This issue will, to an extent, always affect SBI analysis when applied to real observations. 
    To test it, we repeat the analysis, changing the observed {\it Planck} data vector for a synthetic observation. The new data vector is obtained by running {\tt CAMB} at a fiducial cosmology, adding noise from the noise model described in~\cref{sec:simulator}, and then adding extra Gaussian noise at small scales $\ell > 1000$. We choose this multipole range because these are the scales at which Silk damping~\cite{hu1997physics} is the dominant effect. Therefore this artificial systematic could be interpreted as some unknown physics associated with Silk damping. Given that this extra noise has been added to any of the training simulations, our observed data is different from any of the simulations our DELFI algorithm has been trained on. 
    
    The results of the analysis are shown in~\cref{fig:results_noisy}. Using no explicit compression, we get consistent results, but we can see that when we do not marginalise over the weights with SWA, we get biased posteriors in some parameters. 
    %, in this case, especially in $\log A$. 
    In this case, because we know the true parameters, we can calculate the excess probability (EP) of the true parameters, as described in~\ref{sec:validation}. In this case, we get $EP =  0.05$ without weight marginalisation, and $EP = 0.18$ when marginalising over weights, showing that weight marginalisation does improve our results. In~\ref{sec:validation_generalization}, we repeat this for all simulations in the validation set and find that indeed the non weight-marginalised case is overconfident, whereas with weight marginalisation we can get good constraints by adjusting the scale hyperparameter.
    
    When we use a regression network for compression, the results without weight marginalisation show very clear and dramatic biases, with an excess probability of $EP \sim 2 \cdot 10^{-3}$. This shows that neural compression can lead to dangerous biases when the observed data is different from the simulations.
    This is very much in line with the fact that simple neural networks generally do not handle covariate shift very well, since they may include computations that involve combining irrelevant variables in such a way that a distribution shift can lead to drastic changes in outcomes.
    Weight marginalising greatly improves the reliability of these results, at the expense of increasing the error bars $EP \sim 0.13$. This is expected, given the better generalization properties of BNNs. Therefore, unless we are fully confident that our simulations contain all the observational effects that affect the data, we strongly recommend using weight marginalisation, to avoid biased results. \ref{sec:validation_generalization} repeats this analysis for several validation simulations and again shows biased contours when using compression if we do not marginalise over weights. Thus, we see how in both cases, the generalization properties of BNNs mean that SWA greatly increases the reliability of our SBI analysis when simulations do not perfectly match the data.
    
    \section{Conclusions}
    \label{sec:conclusions}
    
    In this work, we have shown how to address some difficulties encountered in DELFI analyses. 
    We have shown that, in the case of limited simulations, 
    we get larger posterior distributions, and therefore lose constraining power, whether we use massive data compression or not.
    We also show how using DELFI without explicit compression leads to comparable posteriors. 
    In either case, marginalisation of the neural network parameters prevents overfitting, and increases the reliability of the posteriors, at the expense of slightly less confident posteriors. We show how to do this using {\tt cosmoSWAG}, the first application of Stochastic Weight Averaging to cosmology. Finally, we show that weight marginalisation is even more important in the case of simulations that do not perfectly capture the physics of the data. In that case, DELFI without weight marginalisation can lead to strongly biased results. Therefore, in the likely scenario of imperfect simulations, we recommend adding weight marginalisation to your SBI analysis to increase the reliability of the posteriors.

    \section*{Acknowedgements}

    We organize the referees at the ML4Astro Machine Learning for Astrophysics Workshop at the Thirty-ninth International Conference on Machine Learning (ICML 2022), for comments and feedback in previous versions of this work.
    PL acknowledges support by the UK STFC grant
    ST/T000473/1.

    \section*{References}

    \bibliography{main}
    \bibliographystyle{iopart-num}    
    
    %%%%%%%%%%%%%%%%%%%%%%%%%%%%%%%%%%%%%%%%%%%%%%%%%%%%%%%%%%%%%%%%%%%%%%%%%%%%%%%
    %%%%%%%%%%%%%%%%%%%%%%%%%%%%%%%%%%%%%%%%%%%%%%%%%%%%%%%%%%%%%%%%%%%%%%%%%%%%%%%
    % APPENDIX
    %%%%%%%%%%%%%%%%%%%%%%%%%%%%%%%%%%%%%%%%%%%%%%%%%%%%%%%%%%%%%%%%%%%%%%%%%%%%%%%
    %%%%%%%%%%%%%%%%%%%%%%%%%%%%%%%%%%%%%%%%%%%%%%%%%%%%%%%%%%%%%%%%%%%%%%%%%%%%%%%
    \newpage
    \appendix
    \onecolumn
    \section{Prior}
    \label{sec:prior}
    
    Table~\cref{tab:prior} shows the prior distributions for the cosmological parameters used to generate our suite of simulations. In this table, $H_0$ is the Hubble parameter in ${\rm km \ s^{-1} \ Mpc^{-1}}$, $\Omega_b$ and $\Omega_c$ are the energy density of baryons and cold dark matter respectively, $h$ is the reduced Hubble parameter ($h = H_0 [{\rm km \ s^{-1} \ Mpc^{-1}}] / 100$), and $A_s$ and $n_s$ are the amplitude and tilt of the primordial power spectrum. This choice of parameter space is the one typically adopted by CMB analyses~\cite{aghanim2020planck}. 
    
    Note that our simulations assume a flat $\Lambda {\rm CDM}$ cosmology, and fix the optical depth to reionization to $\tau_\mathrm{re} = 0.06$, and the {\it Planck} calibration parameter to $A_{\rm Planck} = 1$.
    
    \begin{table}[t]
        \caption{The prior distribution used to generate simulations.}
        \label{tab:prior}
        \vskip 0.15in
        \begin{center}
        \begin{small}
        \begin{sc}
        \begin{tabular}{lc}
        \toprule
        Parameter & Prior \\
        \midrule
        $H_0$  &  $ \mathcal{U} (50, 90)$ \\
        $\Omega_b h^2$  &  $ \mathcal{U} (0.01, 0.05)$ \\
        $\Omega_c h^2$  &  $ \mathcal{U} (0.01, 0.5)$ \\
        $\log(10^{10} A_\mathrm{s}) $  &  $ \mathcal{U} (1.5, 3.5)$ \\
        $n_s$  &  $ \mathcal{U} (0.8, 1)$ \\
        \bottomrule
        \end{tabular}
        \end{sc}
        \end{small}
        \end{center}
        \vskip -0.1in
    \end{table}
    
    \section{Mixture Density Network}
    \label{sec:MDN}
    
    In this section, we describe the Mixture Density Network (MDN), used for compression-free DELFI introduced in~\cref{sec:analysis}. Our MDN is simply a neural network, taking as inputs the data, and outputting $n_{\rm out}$ outputs, where
    
    \begin{equation}
        n_{\rm out} = \left[n_{\rm \theta} + n_{\rm \theta} \cdot {(n_{\rm \theta} + 1) \over 2} + 1 \right] + n_{\rm comp}.
    \end{equation}
    with $n_{\rm \theta}$ the number of parameters in the parameter space (in this case 5), and $n_{\rm comp}$ is the number of components in our MDN (which we set to 3). In this equation, the first term inside square brackets represents the means of the Gaussian distributions $\mu$, the second term are the non-zero elements of the lower triangular matrix obtained from a Cholesky decomposition of the covariance matrix $\Sigma$, and the last one is the weight of that component of the Mixture Density Network $\alpha$. Therefore, this neural network directly gives us an estimate of the posterior distribution as: 
    
    \begin{equation}
        P(\theta | D) = \sum_{i=1}^{n_{\rm comp}} \alpha_i (D) \cdot N(\theta | \mu_i(D), \Sigma_i(D)),
    \end{equation}
    where $\theta$ and $D$ and the parameters and data respectively.

    \section{Validation}
    \label{sec:validation}
    
    When we know the true parameter values, as is the case in the analysis using a synthetic data vector of~\cref{sec:generalization}, we can calculate the excess probability of the true parameter values. We do this by estimating the probability of a large number of samples from our posterior and calculating the percentage of those samples with a probability smaller than the probability of the true parameters. Therefore, a small excess probability means that the true parameters are very unlikely, and our SBI analysis is very likely to be biased. 
    
    To check if our SBI analysis is biased, we need to repeat this excess probability calculation for numerous validation simulations, and check if the distribution of excess probabilities is uniform~\cite{levasseur2017uncertainties, hermans2021averting}. Equivalently, we can calculate the coverage probability, as the cumulative distribution function of the expected probabilities, and then compare it with the expected coverage probability. 
    
    \subsection{Validation of the default analysis}
    \label{sec:validation_default}
    
    \begin{figure*}[ht!]
        %\vskip 0.2in
        \begin{center}
        \centerline{\includegraphics[width=0.99\columnwidth]{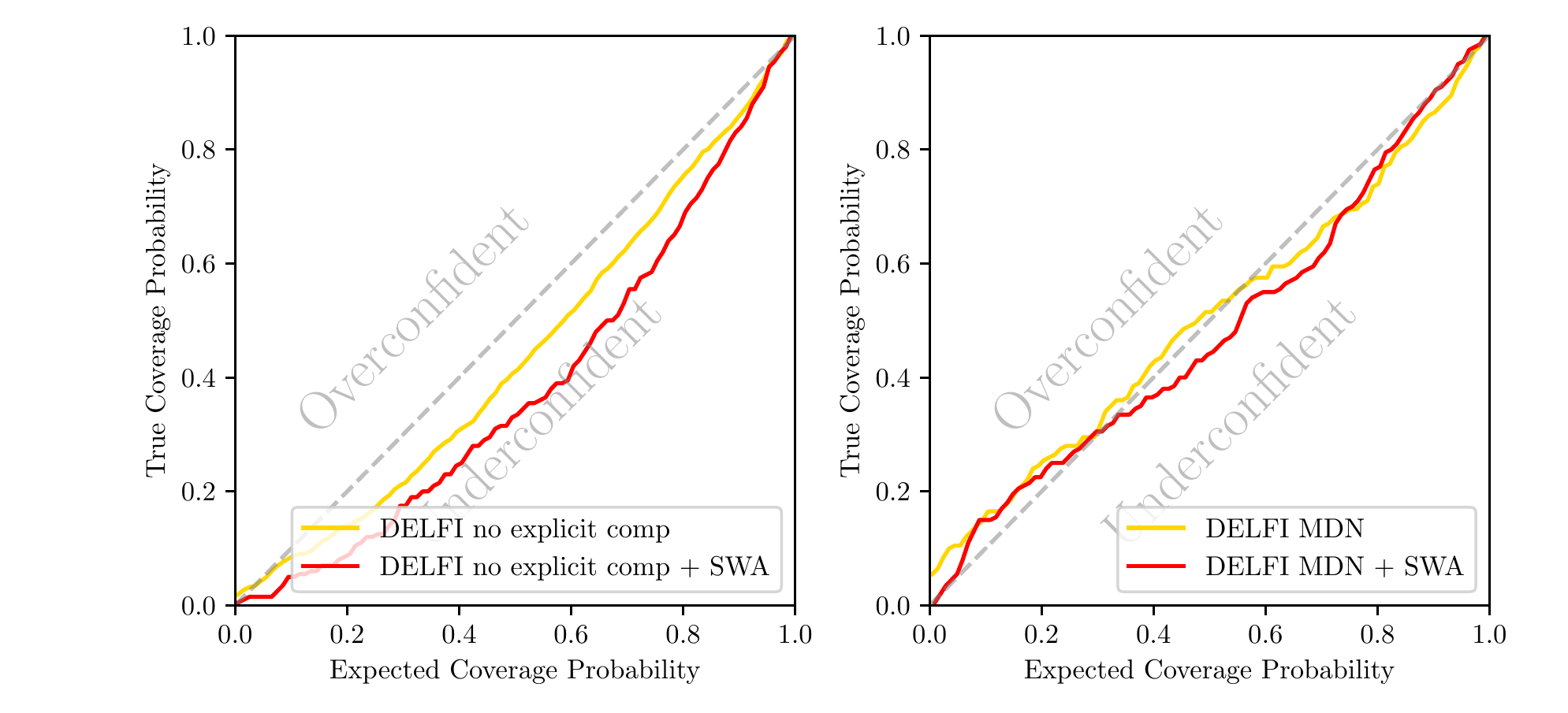}}
       %\vspace{-0.3cm}
        \caption{
            Validation of the default analysis shown in~\cref{fig:results}.
            }
        \label{fig:validation_default}
        \end{center}
        %\vskip -0.2in
    \end{figure*}
    
    We first apply this validation test to the analysis of~\cref{sec:likelihood_analysis}. The results are shown in~\cref{fig:validation_default}. As discussed in the main text, the no explicit compression case gets good results before weight marginalisation, and in fact, marginalisation leads to very underconfident posteriors even when we use a small scale hyperparameter. This is because the weight marginalisation case uses the average of the weights over the SWA training. On the other hand, the compression case gets good posteriors, even when before we use marginalisation.
    
    \begin{figure*}[ht!]
        %\vskip 0.2in
        \begin{center}
        \centerline{\includegraphics[width=0.99\columnwidth]{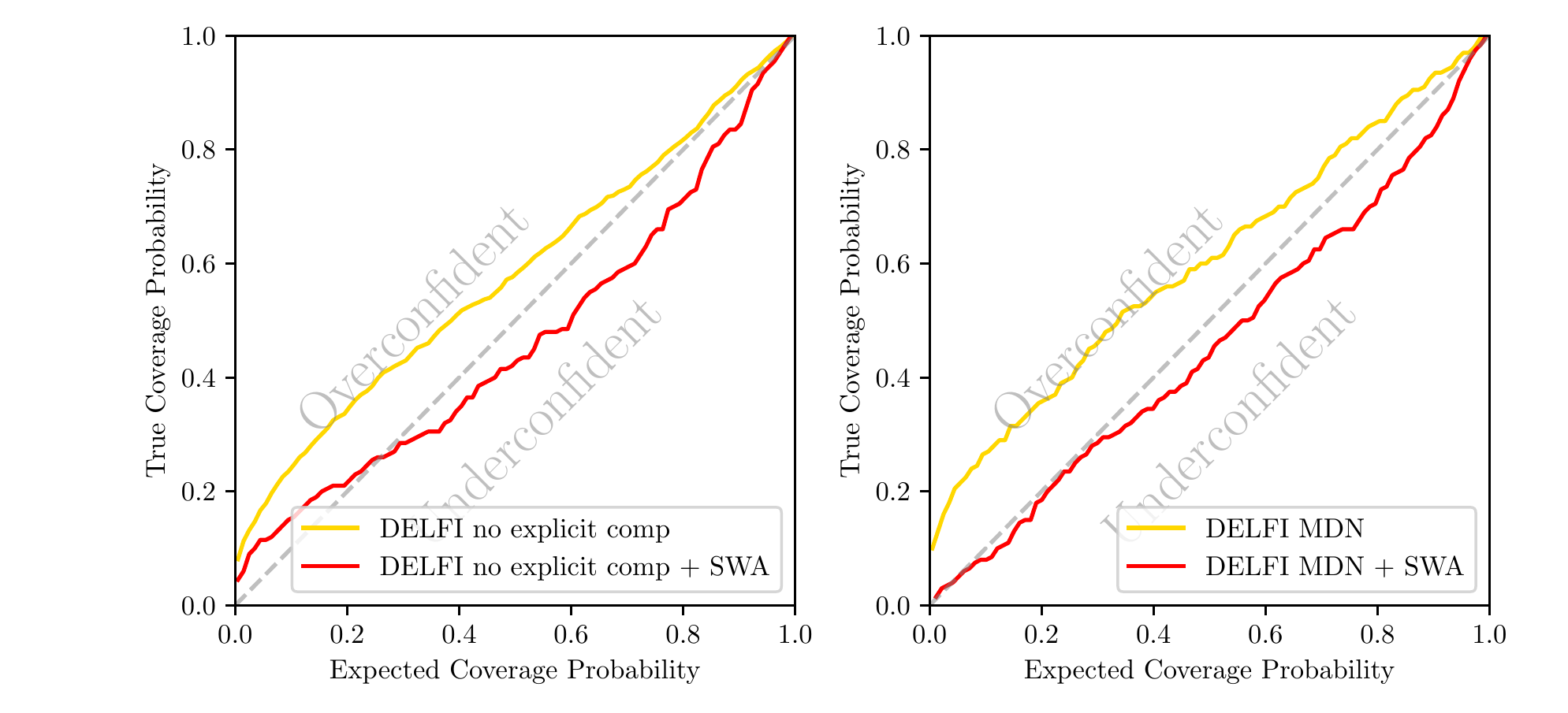}}
        %\vspace{-0.3cm}
        \caption{
            Validation of the generalization analysis shown in~\cref{fig:results_noisy}.
            }
        \label{fig:validation_generalization}
        \end{center}
        %\vskip -0.2in
    \end{figure*}

    \subsection{Validation of the generalization analysis}
    \label{sec:validation_generalization}
    
    We next apply this validation test to the analysis of~\cref{sec:generalization}. For that, we add the extra noise at $\ell > 1000$ for all the simulations in the validation set. The results are shown in~\cref{fig:validation_generalization}. In this case, adding weight marginalisation allows us to get posteriors of the correct size, with and without data compression.
    
    \section{Tuning the Scale Hyperparameter}
    \label{sec:scale}
    
    \begin{figure}[ht!]
        \begin{center}
        \centerline{\includegraphics[width=0.6\columnwidth]{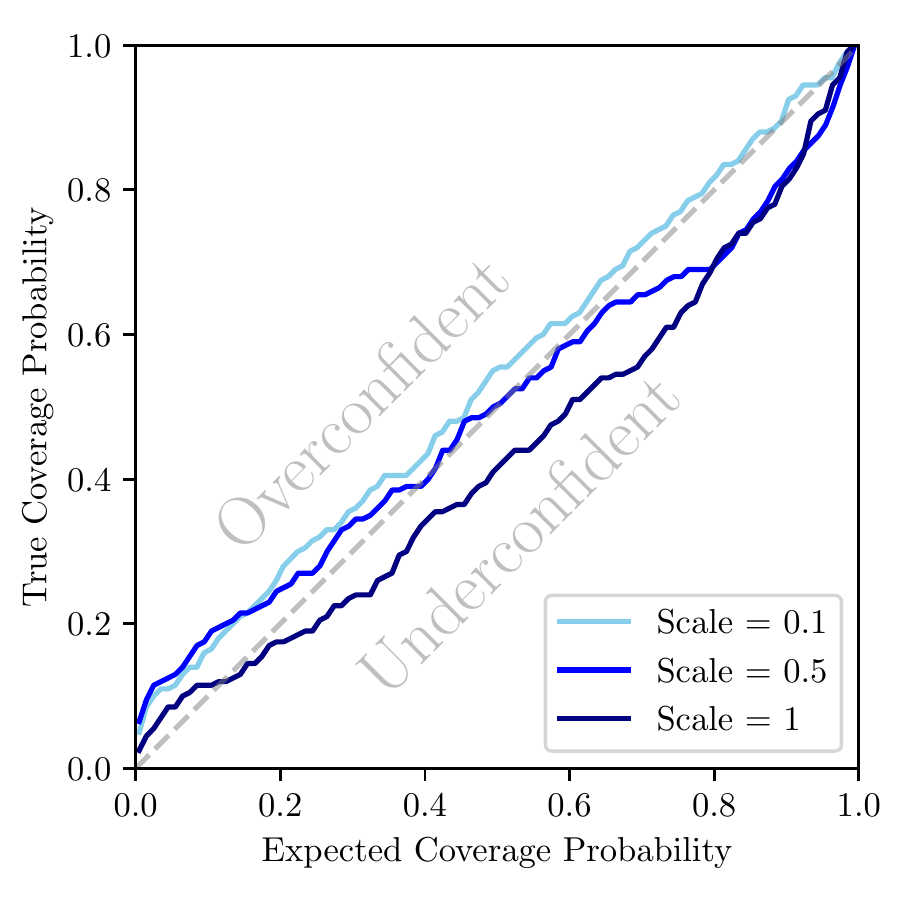}}
        \caption{Illustration of how the SWA scale hyperparameter can be calibrated using the validation test described in~\ref{sec:validation}. A larger scale factor leads to more underconfident posteriors, therefore it can be modified to get as close as possible to the diagonal.} 
        \label{fig:scale}
        \end{center}
    \end{figure}
    
    As described in the main text, the SWA algorithm allows us to rescale the covariance matrix by a scale hyperparameter, to correct for the fact that the covariance matrix can depend on the learning rate used~\cite{maddox2019simple}. In this work, we adjust the scale hyperparameter using the validation test described in~\ref{sec:validation}. More specifically, we adjust the scale so the line in our coverage probability plots gets as close as possible to the diagonal, erring on the side of underconfident posteriors, to avoid biased results. This is illustrated by~\cref{fig:scale}, which shows this calibration performed for the DELFI with no explicit compression analysis applied to noisy data vectors of~\cref{sec:generalization}. 
    In the figure, we see that a scale of $0.1$ leads to overconfident posteriors, and even a scale of $0.5$ is too overconfident. When we raise the scale to $1$, we find that the line is predominantly under the diagonal, therefore we set the hyperparameter to that value. The advantage of tuning this hyperparameter is that it does not require retraining the network, and therefore different values can be tested fast. 
    
    In this work, we used approximate criteria consisting of looking at the coverage probability plots. In future work, we will explore more exact tests and algorithmic tuning of the scale hyperparameter.
    
    \section{Dependence on the number of simulations}
    \label{sec:number}
    
    \begin{figure*}[ht!]
        \begin{center}
        \centerline{\includegraphics[width=0.95\columnwidth]{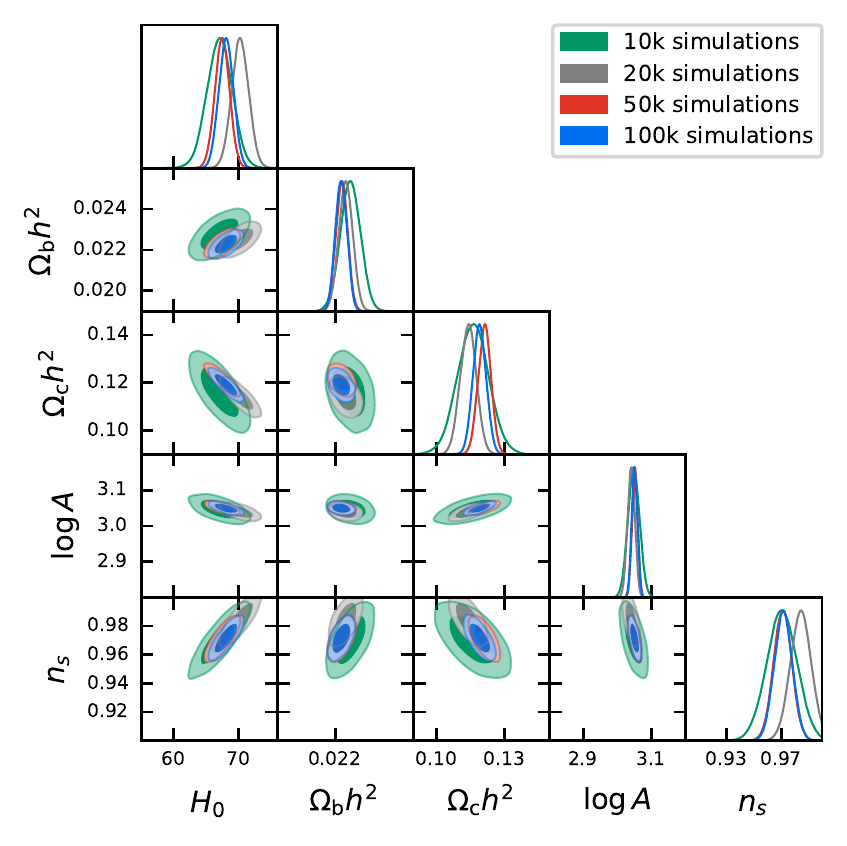}}
        \caption{
            Dependence on the number of simulations. The different contours show the results from the DELFI + MDN analysis when we vary the number of available simulations. 
            }
        \label{fig:number_sims}
        \end{center}
    \end{figure*}
    
    In this section, we explore the dependence of our results on the number of available simulations on our posterior inference. The number of simulations is a key parameter in the analysis, as it affects the accuracy of the posterior inference. Simulations are computationally expensive to run in a lot of contexts, and therefore it is important to understand how the number of simulations affects the results. 

    We repeat the analysis of~\cref{sec:delfi_no_compression} for different numbers of simulations, and show the results in~\cref{fig:number_sims}. We see that the size of the inferred posterior gets smaller as the number of simulations increases. By the team we reach $50.000$ simulations, the posterior size converges, and does not significantly change as more simulations are added. Therefore, we can conclude that $100.000$ is a large enough number for our analysis. 
    
    \end{document}